\documentclass[aps,prl,twocolumn,amsfonts,amssymb,amsmath,floats,floatfix,showpacs,preprintnumbers,superscriptaddress]{revtex4-1}

\usepackage{graphicx}
\usepackage{color}
\usepackage{dcolumn}
\usepackage{bm}
\newcommand{\op}[1]{\ensuremath{\vec{#1}}}

\begin{document}

\title{Layer-by-layer entangled spin-orbital texture of the topological surface state in Bi$_2$Se$_3$}

\author{Z.-H. Zhu}
%\email{zhzhu@physics.ubc.ca}
\affiliation{Department of Physics {\rm {\&}} Astronomy, University of British Columbia, Vancouver, British Columbia V6T\,1Z1, Canada}
\author{C.N. Veenstra}
\affiliation{Department of Physics {\rm {\&}} Astronomy, University of British Columbia, Vancouver, British Columbia V6T\,1Z1, Canada}
\author{G. Levy}
\affiliation{Department of Physics {\rm {\&}} Astronomy, University of British Columbia, Vancouver, British Columbia V6T\,1Z1, Canada}
\author{A. Ubaldini}
\affiliation{D\'{e}partment\,de\,Physique\,de\,la\,Mati\`{e}re\,Condens\'{e}e, Universit\'{e} de Gen\`{e}ve, CH-1211 Gen\`{e}ve 4, Switzerland}
\author{P. Syers}
\affiliation{CNAM, Department of Physics, University of Maryland, College Park, Maryland 20742, USA}
\author{N.P. Butch}
\affiliation{CNAM, Department of Physics, University of Maryland, College Park, Maryland 20742, USA}
\author{J. Paglione}
\affiliation{CNAM, Department of Physics, University of Maryland, College Park, Maryland 20742, USA}
\author{M.W. Haverkort}
\affiliation{Max Planck Institute for Solid State Research, Heisenbergstra\ss{}e 1, D-70569 Stuttgart Germany }
\affiliation{Quantum Matter Institute, University of British Columbia, Vancouver, British Columbia V6T\,1Z4, Canada}
\author{I.S. Elfimov}
\affiliation{Department of Physics {\rm {\&}} Astronomy, University of British Columbia, Vancouver, British Columbia V6T\,1Z1, Canada}
\affiliation{Quantum Matter Institute, University of British Columbia, Vancouver, British Columbia V6T\,1Z4, Canada}
\author{A. Damascelli}
\email{damascelli@physics.ubc.ca}
\affiliation{Department of Physics {\rm {\&}} Astronomy, University of British Columbia, Vancouver, British Columbia V6T\,1Z1, Canada}
\affiliation{Quantum Matter Institute, University of British Columbia, Vancouver, British Columbia V6T\,1Z4, Canada}

\date{\today}

\begin{abstract}
We study Bi$_2$Se$_3$ by polarization-dependent angle-resolved photoemission spectroscopy (ARPES) and density-functional theory slab calculations. We find that the surface state Dirac fermions are characterized by a {\it layer-dependent} entangled spin-orbital texture, which becomes apparent through quantum interference effects. This explains the discrepancy between the spin polarization from spin-resovled ARPES -- ranging from 20 to 85\% -- and the 100\% value assumed in phenomenological models. It also suggests a way to probe the intrinsic spin texture of topological insulators, and to continuously manipulate the spin polarization of photoelectrons and photocurrents all the way from 0 to $\pm$100\% by an appropriate choice of photon energy, linear polarization, and angle of incidence.
\end{abstract}


\maketitle

Topological insulators (TIs) define a new state of matter in which strong spin-orbit interaction (SOI) leads to the emergence of a metallic topological surface state (TSS) formed by spin-nondegenerate Dirac fermions \cite{Kane:z2, zhang:QSH, Fu:2007, Moore:z2, Qi:2008, Hasan:2010PRM}. To capture the physics of TIs, a spin-momentum locking with 100\% spin polarization is usually assumed for the TSS in time-reversal invariant models \cite{Fu:2007, Moore:z2, Qi:2008}. The successful realization of topological insulating behavior in quantum wells \cite{Bernevig, Konig} and crystalline materials such as Bi$_2$Se$_3$ \cite{Zhang:2009BeSe, Xia:2009BiSe, Chen:2010BiSe}  brings us closer to the practical implementation of theoretical concepts built upon novel topological properties. However, the large discrepancy in the degree of TSS spin polarization determined for Bi$_2$Se$_3$ by spin-resolved ARPES (angle-resolved photoemission spectroscopy) -- ranging from 20 to 85\% \cite{Hsieh:2009BiSe, Souma:spin, Xu:spin, Pan:spin, Jozwiak} -- challenges the hypothesis of a 100\% spin polarization for real TIs. First principle density-functional theory (DFT) also indicates that the TSS spin polarization in members of the Bi$_2$X$_3$ material family (X=Se, Te) can be substantially reduced from 100\%  \cite{Yazyev:dft,Guo:dft}. Based on general symmetry arguments, it was shown that the spin polarization direction of photoelectrons in spin-resolved ARPES can be very different from that of the TSS wavefunction \cite{park:spin}. However, the role played by the intrinsic properties of the TSS wavefunction in defining the highest spin polarization that could be achieved, for instance in d.c. and photoinduced electrical currents, has remained elusive.

We report here that the TSS many-layer-deep extension into the material's bulk -- in concert with strong SOI -- gives rise to a layer-dependent, entangled spin-orbital texture of the Dirac fermions in Bi$_2$Se$_3$. A remarkable consequence, specifically exploited in this study, is that one can gain exquisite sensitivity to the internal structure of the TSS wavefunction, $\Psi_{\mathrm{TSS}}$, via quantum interference effects in ARPES. The spin-orbital texture is captured directly in the linear-polarization dependence of the ARPES intensity maps in momentum space, and can be fully resolved with the aid of ab-initio DFT slab-calculations. This has also major consequences in the interpretation of spin-resolved ARPES results, explicitly solving the puzzle of the TSS spin polarization, and suggesting how 100\% spin polarization of photoelectrons and photocurrents can be achieved and manipulated in TI devices by using linearly polarized light.

\begin{figure*}[t!]
\includegraphics[width=1\linewidth]{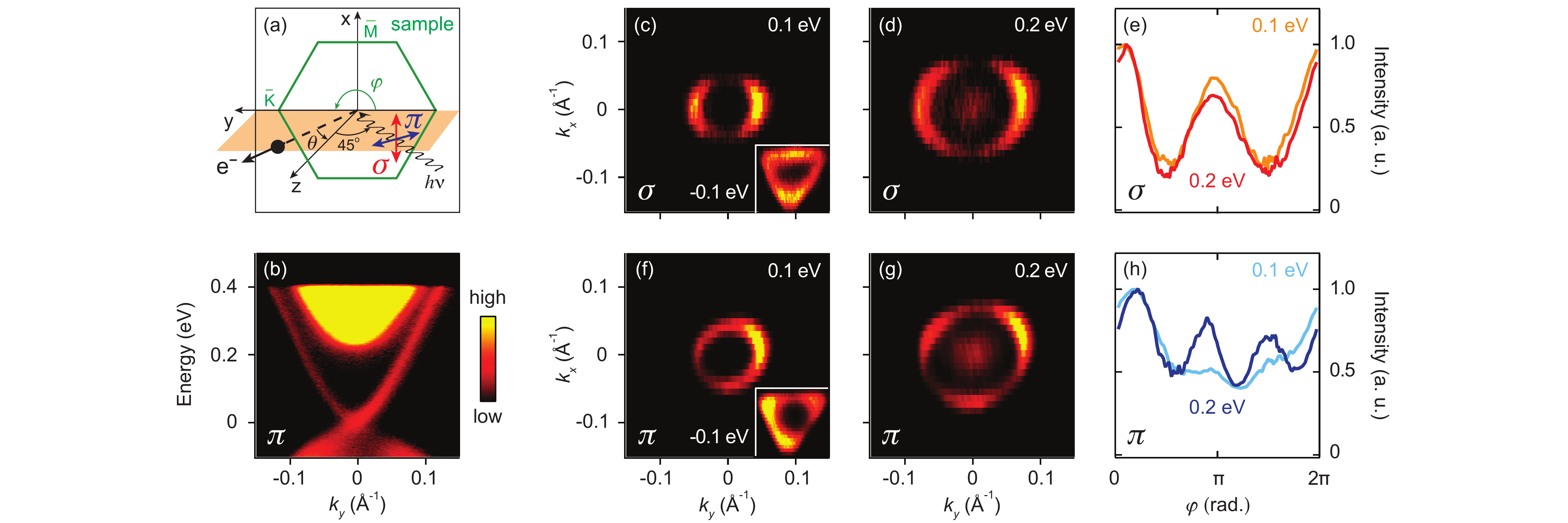}
\caption{\label{fig:ARPES} (color online). (a) Schematics of the experimental geometry, with $\pi$ (horizontal) and $\sigma$ (vertical) linear polarizations, and horizontal photoelectron emission plane. (b) ARPES dispersion measured along $\bar{\mathrm{K}}\!-\!\bar{\mathrm{\Gamma}}\!-\!\bar{\mathrm{K}}$ with $\pi$ polarization; the zero of energy has been set at the Dirac point (DP) for convenience. (c), (d) Constant energy ARPES maps from above (0.1 and 0.2\,eV) and below (-0.1\,eV, inset) the DP, measured with $\sigma$ polarization; (f), (g) same for $\pi$ polarization. (e), (h) Normalized variation of the $\sigma$- (e) and $\pi$-polarization (h) ARPES intensity, along the Dirac contours, plotted versus the in-plane angle $\varphi$.}
\end{figure*}

We start our discussion from the Bi$_2$Se$_3$ ARPES results in Fig.\,\ref{fig:ARPES}, measured with $\sigma$ and $\pi$ linearly-polarized 21.2\,eV photons \cite{supplementary, Zhu:RB}. Based on the experimental geometry [Fig.\,\ref{fig:ARPES}(a)] and photoemission selection rules, $\sigma$-polarization probes the in-plane $p_x$ and $p_y$ orbitals, whereas $\pi$-polarization a combination of both in-plane and out-of plane ($p_z$) orbitals: the 80\% overall intensity reduction observed by switching from $\pi$- to $\sigma$-polarization indicates that the TSS has a dominant $p_z$ character. 
As for the evolution of the ARPES intensity around the Dirac cone, in $\sigma$-polarization [Fig.\,\ref{fig:ARPES}(c)-(e)] we observe a twofold pattern at both 0.1 and 0.2\,eV above the Dirac point (DP), consistent with a previous report \cite{Cao:s}, although somewhat asymmetric with respect to the $k_y\!=\!0$ plane [see in particular Fig.\,\ref{fig:ARPES}(e)]; this suggests a tangential alignment of the in-plane $p_{x,y}$ orbitals with respect to the Dirac constant-energy contours. Conversely in $\pi$-polarization [Fig.\,\ref{fig:ARPES}(f)-(h)] we observe a strongly asymmetric pattern at 0.1\,eV above the DP, which evolves into a triangular pattern while still retaining some asymmetry at 0.2\,eV above the DP; this is in stark contrast with the uniform distribution of intensity along the Dirac contour expected for the dominant $p_z$ orbitals. Finally, at $-0.1$\,eV below the DP, a triangular pattern is observed for both polarizations [see insets of Figs.\,\ref{fig:ARPES}(c) and\,\ref{fig:ARPES}(f ))].

The asymmetry in ARPES intensity between $\pm\mathbf{k_\parallel}$ is particularly evident in $\pi$-polarization at 0.1\,eV in Fig.\,\ref{fig:ARPES}(f) and in the band dispersion of Fig.\,\ref{fig:ARPES}(b). This finding, which might seem in conflict with the time-reversal invariance of the TSS, provides fundamental clues on the structure of $\Psi_{\mathrm{TSS}}$. Time-reversal invariance requires the state at $+\mathbf{k}$ with (pseudo) spin up to be degenerate with the state at $-\mathbf{k}$ with (pseudo) spin down, i.e. to have the same real-orbital occupation numbers. 
\begin{figure}[b!]
\includegraphics[width=1\linewidth]{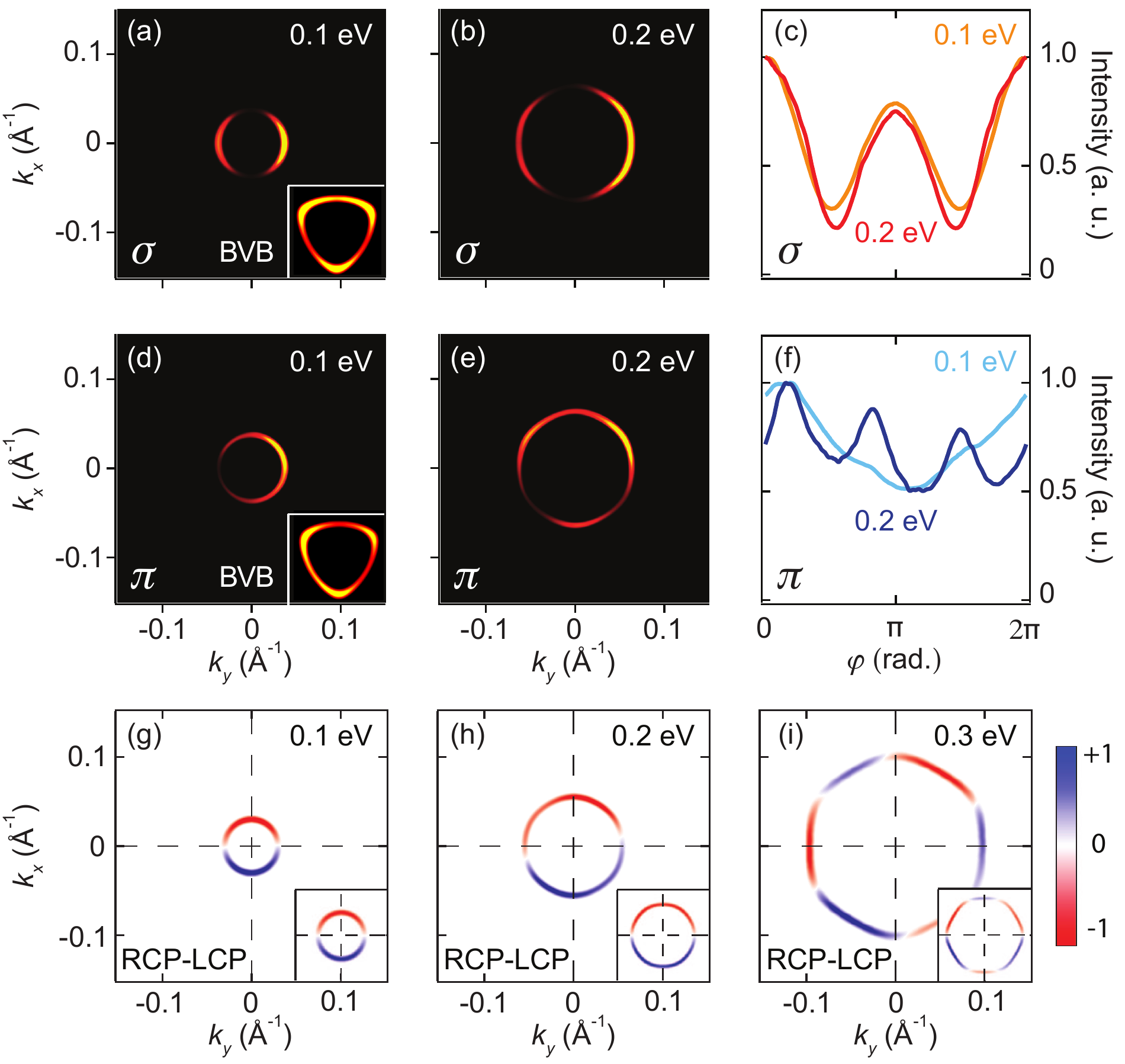}
\caption{\label{fig:TBPHE} (color online). (a),(b) Calculated constant-energy $\sigma$-polarization ARPES maps for TSS (0.1 and 0.2\,eV \cite{renormalization}) and bulk valence band (BVB, -0.1\,eV in the inset); (c) corresponding variation of the ARPES intensity versus the in-plane angle $\varphi$. (d)-(f) Same data as in (a)-(c), but now for $\pi$-polarization. (g)-(i) Calculated  constant-energy circular dichroism ARPES patterns at 0.1, 0.2, and 0.3\,eV above the DP; insets: patterns obtained by rotating the sample by $90\,^{\circ}$ about the normal.} 
\end{figure}
\begin{figure*}[t!]
\includegraphics[width=0.98\linewidth]{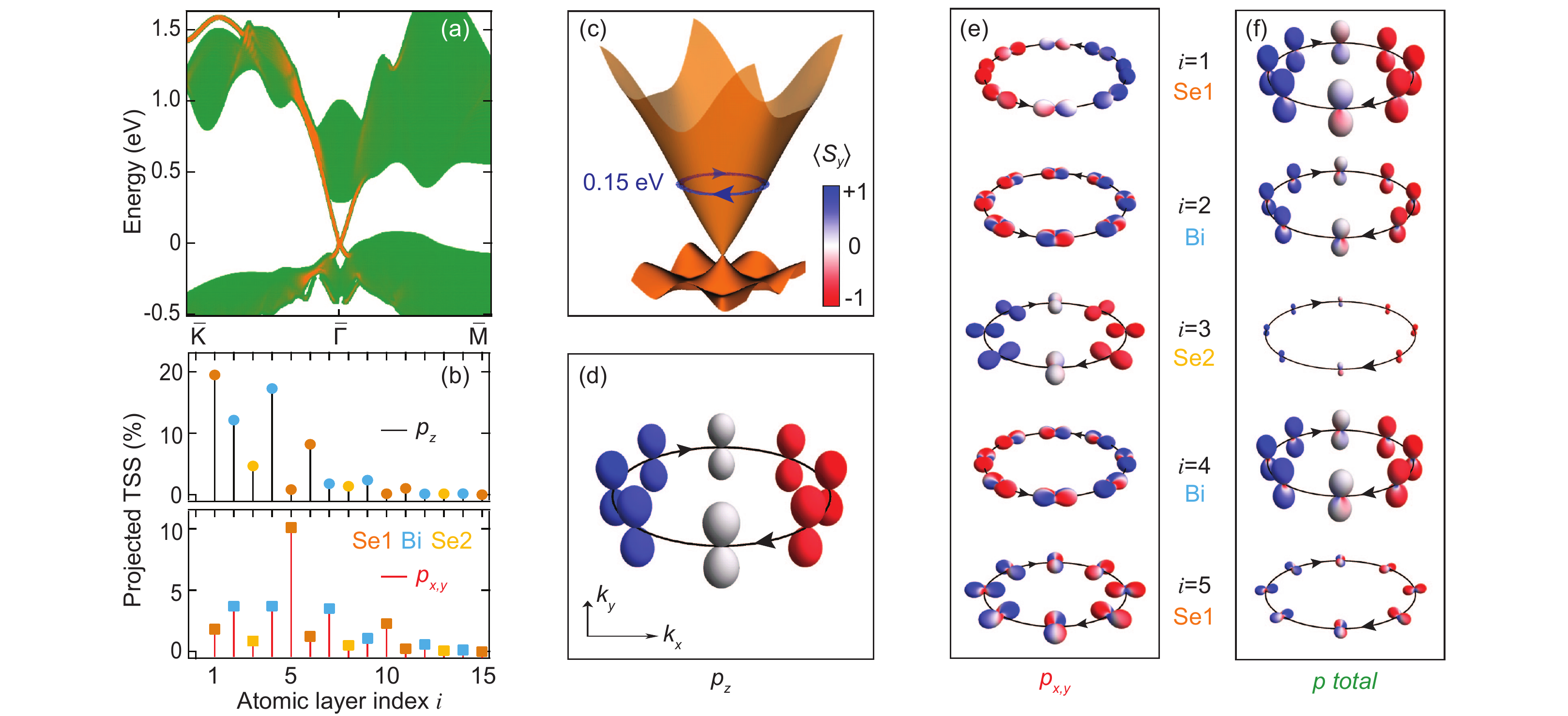}
\caption{\label{fig:TBORB} (color online). (a) Electronic dispersion from our 250-layer-slab DFT model \cite{supplementary}, with TSS in orange and bulk states in green \cite{SS}. (b) Percentage contribution of $p_z$ and ${p_{x,y}}$ orbitals to $\Psi_{\mathrm{TSS}}$ at 0.15\,eV above the DP, resolved layer-by-layer, for the top 15 atomic layers. (d)-(f) Layer- and orbital-projected charge density along the 0.15\,eV $k$-space contour indicated in (c); their surfaces are defined by $r(\theta,\phi)=\sum_{\tau,\tau'} Z_{\tau}(\theta,\phi) Z_{\tau'}(\theta,\phi) \langle a^{\dag}_{i,\tau,k}a^{\,}_{i,\tau',k} \rangle$, where $i$ and $\tau$ are layer and orbital basis indexes and $Z$ the cubic harmonics, and colored according to the expectation value of the $S_y$ operator. The {\it total} layer-resolved TSS texture (f) is obtained by adding all $p$ orbital contributions according to their relative, layer-dependent weight from panel (b).} 
\end{figure*}
This so-called Kramers degeneracy, together with the ARPES selection rules for linearly polarized light, forbids intensity patterns which are different at $\pm\mathbf{k}$. We emphasize here that this restriction can be rephrased in terms of purely in-plane momentum coordinates, i.e. $\pm\mathbf{k_\parallel}$, only for a perfect 2-dimensional TSS with a delta-function-like density, for which $k_z$ plays no role. Thus the observation of an imbalance in ARPES intensity at $\pm\mathbf{k_\parallel}$, together with the established time-reversal invariance of TIs,  necessarily implies that  $\Psi_{\mathrm{TSS}}$ must have a finite extent -- albeit not a dispersion \cite{Bianchi:2D} -- along the third dimension. While details will become clear when discussing our DFT results in Fig.\,\ref{fig:TBORB}, we anticipate that this -- together with SOI -- leads to a complex layer-dependent spin-orbital entanglement in Bi$_2$Se$_3$, which becomes apparent in ARPES through photoelectron interference.

By performing ARPES intensity calculations \cite{supplementary,Bansil} for TSS and bulk wavefunctions from our DFT slab-calculations, we accurately reproduce the data. As shown in Fig.\,\ref{fig:TBPHE}(a)-(f), we obtain very different intensities at $\pm\mathbf{k_\parallel}$ in excellent agreement with the results for both $\sigma$ and $\pi$ polarizations. Specifically, we reproduce the quasi-twofold pattern in $\sigma$ polarization, stemming from the spatial configuration of $p_{x,y}$ orbitals [Fig.\,\ref{fig:TBPHE}(a) and (b)]; the quasi-threefold pattern away from the DP [Fig.\,\ref{fig:TBPHE}(e)], which originates from the hybridization between TSS and bulk states \cite{Ong:bulk,kz}; and also the triangular patterns at -0.1\,eV [insets of Fig.\,\ref{fig:TBPHE}(a) and (d)]. Note that the ARPES intensity visible at the $\bar\Gamma$ point in Fig.\,\ref{fig:ARPES}(d) and (g), but not reproduced by our calculations, originates from the scattering-induced broadening of the bulk conduction band \cite{impurity}. As a final test of the robustness of our DFT analysis of $\Psi_{\mathrm{TSS}}$, we have calculated constant-energy circular dichroism ARPES patterns, which are also in excellent agreement with previous studies \cite{wang:CD,park:CD}.  

To gain a microscopic understanding of the properties of $\Psi_{\mathrm{TSS}}$ we present our DFT results for a 250-layer slab of Bi$_2$Se$_3$ \cite{supplementary} in Fig.\,\ref{fig:TBORB}(a), with bulk states in green and TSS in orange. The in- and out-of-plane $p$ orbital  projections in Fig.\,\ref{fig:TBORB}(b) confirm that $\Psi_{\mathrm{TSS}}$ indeed has a large $p_z$ (70\%) character -- although $p_{x,y}$ (30\%) is also significant -- and most importantly that $\Psi_{\mathrm{TSS}}$ extends deep into the solid. Even though the orbital weight decays exponentially with the distance from the surface, as expected for a surface bound-state, $\Psi_{\mathrm{TSS}}$ extends $\sim$2 quintupole layers (QL) below the surface ($\sim2$\,nm), with $\sim$75\% contribution from the 1$^{st}$ QL and $\sim$25\% from the 2$^{nd}$ QL. Note also the interesting layer dependence of the orbital character: while for most layers the main component is the out-of-plane $p_z$, for the 5$^{th}$ the in-plane $p_{x,y}$ is actually dominant.

As a consequence of the relativistic SOI, which directly connects orbital to spin flips via the $l^{\pm} s^{\mp}$ terms of the spin-orbit operator $\mathbf{l}\!\cdot\!\mathbf{s} = l_z s_z\!+\!(l^+ s^-\!+ l^- s^+)/2$, the strongly layer-dependent orbital occupation becomes entangled with the spin polarization of $\Psi_{\mathrm{TSS}}$. To visualize this entanglement, in Fig.\ref{fig:TBORB}(d)-(f) we present the  layer- and orbital-projected charge density along the 0.15\,eV Dirac contour indicated in Fig.\ref{fig:TBORB}(c), colored according to the expectation value of the $S_y$ operator \cite{supplementary}. The $p_z$-projected charge density, being associated with a single orbital, cannot be entangled and has the layer-independent spin helicity shown in Fig.\,\ref{fig:TBORB}(d). In contrast, a strong layer-dependent spin-orbital entanglement is observed for ${p_{x,y}}$ because the eigenstates can be a linear combination of $p_{x,\uparrow}$, $p_{y,\downarrow}$, and similar states, resulting in a complex set of charge-density surfaces.
\begin{figure*}[t!]
\includegraphics[width=0.95\linewidth]{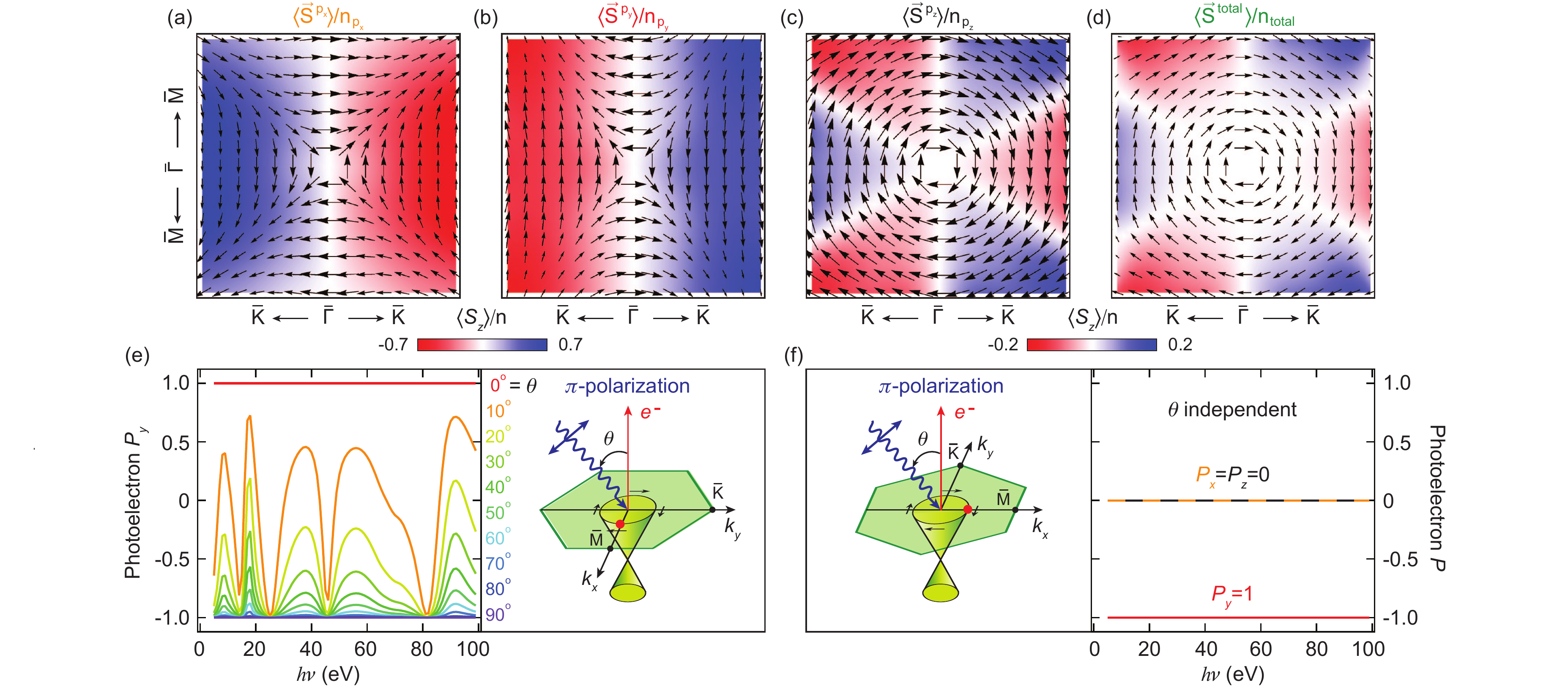}
\caption{\label{fig:TBSpin} (color online). (a)-(d) Spin texture of the Bi$_2$Se$_3$ Dirac cone upper branch (arrows: {\it in-plane}; colors: {\it out-of-plane}), obtained from the expectation value of the layer-integrated, orbital-projected spin operators, normalized to the orbital occupation \cite{supplementary}. Note that (a), (b) and (c), (d) have different color scales but that the arrow scaling is the same; also, moving away from $\bar{\Gamma}$ corresponds to moving in energy away from the DP ($\sim\!0.4$\,eV at maps edge). (e), (f) Prediction for the photoelectron spin polarization ($P$) measured in spin-resolved ARPES as a function of photon energy and incidence angle \cite{supplementary}; two experimental geometries are examined in $\pi$ polarization for the same $k$ point located at 0.15\,eV along $\bar{\Gamma}\!-\!\bar{\mathrm{M}}$ [in (e) only $P_y$ is shown].}
\end{figure*}
These surfaces show two overall spatial configurations oriented tangentially and radially with respect to the Dirac contour, with opposite spin helicity, as seen in Fig.\,\ref{fig:TBORB}(e). In Fig.\,\ref{fig:TBORB}(f) we show the total layer-dependent charge density obtained by adding in- and out-of-plane contributions according to their relative weights in Fig.\,\ref{fig:TBORB}(b); from this it is clear that while the $p_z$ orbitals dominate, the in-plane ${p_{x,y}}$ orbitals lead to a substantial spin-orbital entanglement of the combined $\Psi_{\mathrm{TSS}}$.

This entanglement also leads to complex in- and out-of-plane spin-texture, as shown in Fig.\,\ref{fig:TBSpin}(a)-(d) where the layer-integrated spin patterns of individual and total $p$ orbitals are presented. While for $p_z$ we find the in-plane helical spin texture  expected for the  TSS this is not the case for the $p_{x}$ and $p_y$ orbitals, which exhibit patterns opposite to one another. Combining all contributions [$\langle\op{S}^{\text{total}}\rangle/n_\text{total}$ in Fig.\,\ref{fig:TBSpin}(d)], the TSS out-of-plane spin texture vanishes in the vicinity of the DP; most important, the in-plane spin polarization is reduced from 100\% to 75\% at the DP, and to 60\% at 0.4\,eV above the DP \cite{supplementary}. Note that this is also critically dependent on the relative $p_{x,y}$ orbital content of $\Psi_{\mathrm{TSS}}$, which increases from 25\% to 45\% over the same energy range \cite{supplementary}.

For the discussion of the ARPES intensity \cite{supplementary}, we will here use the approximation $I\!\propto\!|\langle e^{i \mathrm{\bf k} \cdot \mathrm{\bf r}} | \mathrm{\bf A}\!\cdot\!\mathrm{\bf p}  | \Psi_{\mathrm{TSS}} \rangle|^2$, expressed in terms of plane-wave photoelectron final states for simplicity. By writing $\Psi_{\mathrm{TSS}}$ as linear combination of layer-dependent eigenstates, $\Psi_{\mathrm{TSS}}\!=\! \sum_{i,\sigma} \alpha_{i} \psi^{\sigma}_{i,\mathrm{\bf k_{\parallel}}}$ with $i$ and $\sigma$ being layer and spin indexes, the ARPES intensity becomes $I\!\propto\!\sum_{\sigma}| \sum_{i} e^{-i k_{z} z_{i}} \langle e^{i \mathrm{\bf k_{\parallel}}\!\cdot\!\mathrm{\bf r_{\parallel}}} | \mathrm{\bf A} \cdot \mathrm{\bf p}  | \alpha_{i} \psi^{\sigma}_{i,\mathrm{\bf k_{\parallel}}}\rangle |^{2}$. Here the $e^{-i k_{z} z_{i}}$ phase term accounts for the photoelectron optical path difference stemming from the TSS finite extent into the bulk. Because both $e^{-i k_{z} z_{i}}$ and $\psi^{\sigma}_{i,\mathrm{\bf k_{\parallel}}}$ vary from layer to layer [the latter via the relative orbital content as shown in Fig.\,\ref{fig:TBORB}(f)], the photoemission intensity is dominated by interference between the $\psi^{\sigma}_{i,\mathrm{\bf k_{\parallel}}}$ eigenstates, and can in fact be regarded as the Fourier transform of the layer-dependent $\Psi_{\mathrm{TSS}}$. We also note that, because the phase of photoelectrons is defined by additive $k_z$ and $\mathrm{\bf k_{\parallel}}$ contributions, reversing the sign of either $k_z$ or  $\mathrm{\bf k_{\parallel}}$ will change the ARPES intensity, i.e. $I(k_z)\!\neq\!I(-k_z)$ and especially $I(\mathrm{\bf k_{\parallel}})\!\neq\!I(-\mathrm{\bf k_{\parallel}})$ as observed experimentally.

Photoelectron interference also severely affects the spin polarization $P_{x, y, z} = \frac{I^{\uparrow_{x, y, z}} - I^{\downarrow_{x, y, z}}} {{I^{\uparrow_{x, y, z}} + I^{\downarrow_{x, y, z}}}}$ measured in spin-resolved ARPES \cite{supplementary}. This exhibits a strong dependence on photon energy, polarization, and angle of incidence, which in general prevents the straightforward experimental determination of the intrinsic spin-texture of Bi$_2$Se$_3$. While comprehensive results are presented in Fig.\,S2 and S3 \cite{supplementary}, in Fig.\,\ref{fig:TBSpin}(e) and (f) we show as an example the same $k$ point along $\bar{\Gamma}\!-\!\bar{\mathrm{M}}$ measured in two different geometries, probing selectively $p_{y,z}$ (e) and $p_{x,z}$ (f) orbitals. In Fig.\,\ref{fig:TBSpin}(e), because $\langle\op{S}^{p_y}\rangle$ and $\langle\op{S}^{p_z}\rangle$ (the spin polarization of the $p_y$ and $p_z$ orbitals) are antiparallel at this specific $k$ point, $P_y$ varies between $\pm$100\% upon changing $\theta$, and oscillates wildly as a function of photon energy (with the exception of $0\,^{\circ}$ and $90\,^{\circ}$, which probe $p_{y}$ and $p_{z}$ separately). However, if the sample is rotated by $90\,^{\circ}$ as in Fig.\,\ref{fig:TBSpin}(f) $\langle\op{S}^{p_x}\rangle$ becomes parallel to $\langle\op{S}^{p_z}\rangle$ and the measured  $P_{x, y, z}$ are all independent of photon energy and incidence angle, allowing the detection of the intrinsic spin polarization. We note that this behavior is consistent with reported spin-resolved ARPES results \cite{Jozwiak}: for the situation of Fig.\,\ref{fig:TBSpin}(f), $P_y\gtrsim 80$\% was obtained, close to our 100\% expectation; along $\bar{\Gamma}\!-\!\bar{\mathrm{K}}$, $P_y$ was observed to vary from 25\% at $h\nu$ = 36\,eV to $-50$\% at 70\,eV, while we obtain $+20\!\pm\!10\%$ and $-40\!\pm\!15\%$, respectively \cite{supplementary}.

In conclusion, the TSS layer-dependent spin-orbital entanglement is responsible -- via photoelectron interference -- for the apparent time-reversal symmetry breaking in ARPES and the large discrepancy in the estimated TSS spin-polarization from spin-resolved ARPES. This is of critical importance for many applications and fundamental studies of TIs: e.g., the observed $I(\mathrm{\bf k_{\parallel}})\!\neq\!I(-\mathrm{\bf k_{\parallel}})$ provides an explanation for the so-far puzzling result of spin-polarized electrical currents photoinduced by linearly-polarized light \cite{Mciver:2011}, which also is associated with an imbalance in the number of photoelectrons removed at $\pm\mathbf{k_\parallel}$. In addition, exploiting photoelectron interference in spin-resolved ARPES provides a way not only to probe the intrinsic spin texture of TIs, but also -- and most importantly -- to precisely control in- and out-of-plane spin polarization of the photocurrent in spin-resolved ARPES -- all the way from 0 to $\pm$100\% -- by varying energy, polarization, and angle of incidence of the incoming photons.

We gratefully acknowledge M. Franz, G.A. Sawatzky, and H. Guo for discussions. This work was supported by the Max Planck - UBC Centre for Quantum Materials, the Killam, Alfred P. Sloan, Alexander von Humboldt, and NSERC's Steacie Fellowships (A.D.), the Canada Research Chairs Program (A.D.), NSERC, CFI, and CIFAR Quantum Materials. Work at the University of Maryland was supported by NSF-MRSEC (DMR-0520471) and DARPA-MTO award (N66001-09-c-2067).

\bibliography{BiSe_P}

\begin{thebibliography}{34}
\bibitem [{ren()}]{renormalization}%
  \BibitemOpen
  \href@noop {} {\bibinfo  {journal} {Note that the TSS calculations have
  actually been performed for 0.12 and 0.23\,eV, in order to account for the
  DP--conduction-band gap renormalization observed between experiment
  ($\sim$260\,meV) and DFT (300\,meV)}\ }\BibitemShut {NoStop}%
\bibitem [{SS()}]{SS}%
  \BibitemOpen
\bibfield  {journal} {  }\href@noop {} {\bibinfo  {journal}
  {$\Psi_{\mathrm{TSS}}$ is composed of those states that exceed a 20\%
  projection onto the top 5 atomic layers in real space}\ }\BibitemShut
  {NoStop}%
\bibitem [{kz()}]{kz}%
  \BibitemOpen
  \href@noop {} {\bibinfo  {journal} {Note that the threefold pattern is
  associated with the $k_z$ dispersion of valence- and conduction-band bulk
  states}\ }\BibitemShut {NoStop}%
\bibitem [{imp()}]{impurity}%
  \BibitemOpen
\bibfield  {journal} {  }\href@noop {} {\bibinfo  {journal} {In the experiment,
  the bulk conduction band is significantly broadened by disorder-induced
  scattering, which leads to intensity leaking down to lower binding energies
  inside the gap, an effect not accounted for in our calculations. We also note
  that while this same disorder affects to some degree the linewidth and
  electronic filling of the TSS, its main features -- such as dispersion and
  spin-orbital texture close to the Dirac point -- are retained, as a
  manifestation of topological protection \cite{Yazdani:stmBS,
  Kapitulnik:stm}}\ }\BibitemShut {NoStop}%
\end{thebibliography}

\end{document}